\documentclass[aip,rsi,preprint,numerical]{revtex4-1}

\usepackage{amsmath}
\usepackage{amssymb}
\usepackage{graphicx}
\usepackage{psfrag}
\usepackage{sistyle}
\usepackage{color}

\newcommand{\tip}{\mathrm{tip}}
\newcommand{\base}{\mathrm{ref}}
\renewcommand{\d}{\mathrm{d}}

\definecolor{blue}{rgb}{0.3,0.3,1}

\begin{document}

\title{Quadrature phase interferometer for high resolution force spectroscopy}

\author{Pierdomenico Paolino}
\author{Felipe A. Aguilar Sandoval}
\altaffiliation{Universidad de Santiago de Chile, Departamento de F\`\i sica, Avenida Ecuador 3493, Casilla 307, Correo 2, Santiago, Chile}
\author{Ludovic Bellon}
\email[]{ludovic.bellon@ens-lyon.fr}
\affiliation{Universit\'e de Lyon, Laboratoire de physique, ENS Lyon, CNRS, Lyon 69364, France}

\date{\today}

\begin{abstract}
In this article, we present a deflection measurement setup for Atomic Force Microscopy (AFM). It is based on a quadrature phase differential interferometer: we measure the optical path difference between a laser beam reflecting above the cantilever tip and a reference beam reflecting on the static base of the sensor. A design with very low environmental susceptibility and another allowing calibrated measurements on a wide spectral range are described. Both enable a very high resolution (down to $\SI{2.5E-15}{m/\sqrt{Hz}}$), illustrated by thermal noise measurements on AFM cantilevers. They present an excellent long-term stability, and a constant sensitivity independent of the optical phase of the interferometer. A quick review shows that our precision is equaling or out-performing the best results reported in the literature, but for a much larger deflection range, up to a few $\SI{}{\micro m}$.
\end{abstract}

\maketitle

\section{Introduction}

Since its invention by Binnig, Quate and Gerber~\cite{Binnig-1986} more than 20 years ago, Atomic Force Microscopy (AFM) has turned into a mature technique widely spread in many domains (material sciences, biology, nanotechnology...). The detection scheme proposed in the original article was quickly dropped for handier optical methods. In the early ages of AFM, interferometric setups have been investigated by several authors as an option to measure the deflection of the cantilever~\cite{Rugar-1988,denBoef-1989,Rugar-1989,Schonenberger-1989,denBoef-1991,Mulhern-1991}. The introduction of the optical lever technique~\cite{Meyer-1988} however, much simpler to implement and still very sensitive, limited those techniques to a few specialized application where optical access to the cantilever is restricted (cryogenic experiments for instance) or the ultimate precision of the measurement is important~\cite{Mamin-2001,Hoogenboom-2008,Jourdan-2009}. Our setup belongs to this last category: we use thermal noise as a tool to explore nano-mechanics~\cite{Bellon-2008,Paolino-2009-JAP,Paolino-2009-Nanotech,Buchoux-2011,Li-2012,Laurent-2013}, and thus need a negligible detection noise to characterize intrinsic mechanical fluctuations.

We present in this article a few approaches we have explored in order to build a very sensitive detection scheme while limiting as much as possible any slow drift problems. They are all based on differential interferometry~\cite{Schonenberger-1989,denBoef-1991}: 2 beams are produced in the measurement region of the AFM, the first being reflected on a static reference point of the sensor, the second on the free end of the cantilever. The optical path difference between those two beams will thus be directly proportional to the deflection of the cantilever, and knowledge of the wavelength gives a precise calibration of the measurement. The main originality of our work is the use of quadrature phase interferometry~\cite{Bellon-2002-OptCom} in the context of AFM. This approach allows to extend the range of deflection to several $\SI{}{\micro m}$, while retaining the very high resolution (down to $\SI{2.5E-15}{m/\sqrt{Hz}}$ in the latest developments). This unique combination of high precision and large input range paves the way to high resolution force spectroscopy on highly deformable objects.

The article is organized as follows: a first section describes the various configurations of the measurement area, and the quadrature phase interferometric detection scheme common to our realizations. We present in a second part the performances of the apparatus with a few thermal noise measurements, and compare these results with commercial devices, before the concluding remarks.

\section{Experimental setups} \label{Section:Setup}

\begin{figure}
 \begin{center}
	\includegraphics{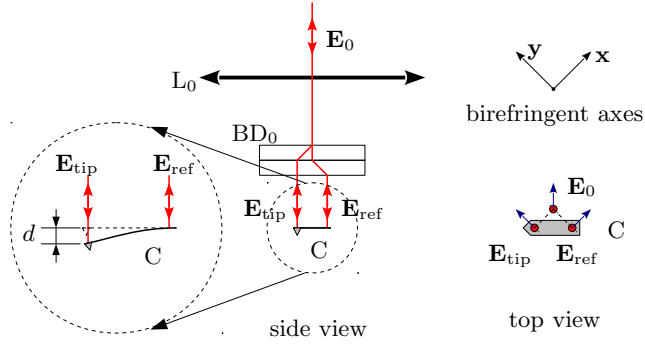}
\end{center}
\caption{Measurement area: the incident beam (electric field $\mathbf{E}_{0}$) is focused on the cantilever C by the lens L$_{0}$. When passing through 2 calcite beam displacers BD$_{0}$ with orthogonal optical axes $\mathbf{x}$ and $\mathbf{y}$, light is split into 2 parallel beams, which are reflected by the base and free end of the sensor (fields $\mathbf{E}_{\base}$ and $\mathbf{E}_{\tip}$). A deflection $d$ of the cantilever modifies the optical path of the second beam by $\delta L=2 d$. On the top view, blue arrows show the direction of polarization of each beam, whereas on the side view, red arrows indicate the direction of propagation of light (to and back from the cantilever).}
\label{Fig-bicalcite}
\end{figure}

Our setup is based on a differential interferometer to measure the deflection of the AFM cantilever: the reference beam is reflected on a static reference point of the sensor, the sensing beam on the free end of the cantilever. The main advantage of this approach is that we decouple the measurement of the deflection from the vibrations of the interferometer: the reference surface is on the sensor itself, and vibrations of the optical setup with respect to the cantilever are cancelled. We will present in the next paragraphs three possible designs of the sensing area, before their common quadrature phase analysis area

\subsection{Interferometer: sensing area}

We developed several strategies to produce the two beams in the measurement region of the AFM. The first one is inspired by reference~\cite{Schonenberger-1989}, where a calcite prism is used. In our setup, as illustrated by Fig. \ref{Fig-bicalcite}, light passes through two calcite beam displacers after the focusing lens. The birefringent axis of the prisms are set perpendicular to each other, hence each polarization of the incident beam is once the ordinary and once the extraordinary ray. Both output beams are then equivalent: their optical path length is equal through the prisms. This setup has 2 advantages: not only the intrinsic optical path is the same on a flat cantilever (canceling the noise due to wavelength fluctuation in the light source), but so does the shift in the focus due to the parallel plates. We usually associate two $\SI{1}{mm}$ thick calcite beam displacers. Each one produces a \SI{100}{\micro m} shift of its extraordinary ray, so the total separation is approximatively \SI{141}{\micro m} and adapted to cantilevers longer than this value. The size of the laser spots is diffraction limited and can be tuned down to a few $\SI{}{\micro m}$ diameter. After reflection, the two beams are merged back together by the birefringent components and can be studied in the analysis area. The optical path difference $\delta L$ is twice the deflection  $d$ of the cantilever:
\begin{equation} \label{eq:deltaL}
\delta L=2 d
\end{equation}

\begin{figure}[t]
 \begin{center}
	\includegraphics{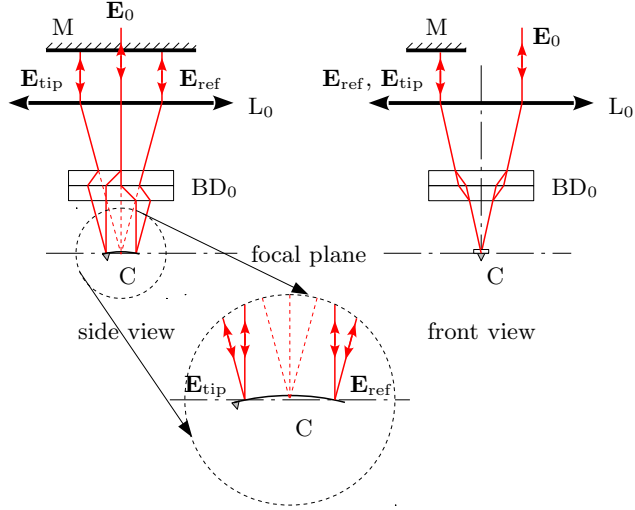} \vspace{-3mm}
\end{center}
\caption{Double path technique: the incident ray (electric field $\mathbf{E}_{0}$) is focused on the sensor by lens L$_{0}$, and split in 2 ($\mathbf{E}_{\base}$ and $\mathbf{E}_{\tip}$) by the calcite prisms (BD$_{0}$). After reflection on the curved surface of the cantilever, both beams exit the lens L$_{0}$ parallel but apart from each other. A perpendicular mirror send them back on their original paths so that they eventually perfectly overlap after a second reflection on the sensor. In practice, we use the 3 dimensions of space: the incident beam $E_{0}$ is slightly off axis on the lens as seen on the front view, so that its reflections $\mathbf{E}_{\base}$ and $\mathbf{E}_{\tip}$ are shifted when exiting L$_{0}$. A single mirror $M$ can be used for both. In this configuration, the optical path difference $\delta L$ is 4 times the deflection $d$.}
\label{Fig-DoublePath}
\end{figure}

One of the problem we had to deal with is the intrinsic curvature of some cantilevers: when those are coated (to enhance reflectivity for instance), the metallic layer may produce internal stresses which lead to a static curvature of the sensor. Strategies to minimize this effect can be used (symmetric coating for example), but the tolerance of common detection systems to this defect is large since most only use a single reflection on the extremity of the cantilever. In our case, the two light beams must overlap in order to record interferences, and the bending of the reflective surface introduces a spatial separation of the 2 polarizations. For instance, a small \ang{2} curvature (typical tolerance of commercially available sensors ) translate into a \SI{2}{mm} separation of the beams back to the \SI{30}{mm} focusing length. The contrast in such a case can decrease to very small values, and depending on the cantilevers we sometime need to address this problem.

When a force acting on the tip is deflecting the cantilever, the previous considerations on the intrinsic curvature applies as well to the induced bending. However, a $\pm\ang{2}$ angular range for a $\SI{300}{\micro m}$ long cantilever translates into a $\pm\SI{7}{\micro m}$ deflection range, much larger than the typical limit of AFM setups.

The simplest solution to deal with curvature limitation is to increase the diameter of the laser beams so that even with such a large separation they still overlap sufficiently to record sharp interferences. Indeed, we can reach a fair contrast on bended cantilevers using a \SI{7}{mm} (instead of our common \SI{2}{mm}) beam, though handling clear apertures for such a large diameter is difficult in our compact setup.

Another way to circumvent the problem is illustrated in Fig. \ref{Fig-DoublePath}. After reflection on the curved surface, the 2 beams corresponding to the 2 polarizations of the birefringent component are parallel when emerging the focusing lens. Indeed, both virtually come from the focal point but with different angles. We place a flat mirror perpendicular to the beams at this point: both are sent back on their original path, and overlap perfectly after a second reflection on the cantilever. This setup also doubles the sensitivity of the measurement: adding a second reflection on the sensor doubles the optical path dependance on the deflection of the cantilever. This configuration can thus be interesting even for flat sensors. However, the tuning of the apparatus is much more complex (the focal distance for instance should be perfectly set as the double pass multiplies any error in this direction). Moreover, another problem can be stressed for uncoated cantilevers: their poorly reflecting surface being used twice, the intensity of the signal is lowered significantly and the overall gain in accuracy is much lower than 2.

\begin{figure}
 \begin{center}
 	\includegraphics{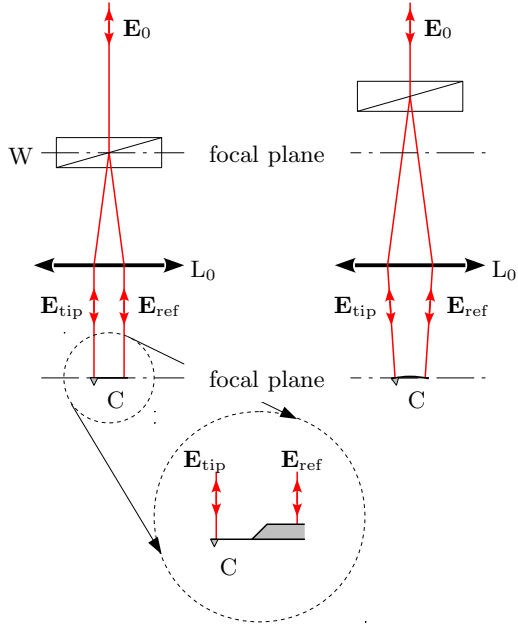}
\end{center}
\caption{Wollaston measurement configuration: the incident ray ($\mathbf{E}_{0}$) is split in its 2 orthogonal polarizations ($\mathbf{E}_{\base}$ and $\mathbf{E}_{\tip}$) by the Wollaston prism, then focused on the sensor by lens L$_{0}$. Tuning the position of birefringent component along the optical axis (right), any small static curvature of the cantilever can be compensated. After reflection, the beams are merged back together and can be analyzed: the optical path difference is twice the deflection. Due to increased distance between the 2 beams (\SI{1}{mm} in our configuration), the reference ($\mathbf{E}_{\base}$) is taken on the chip holding the cantilever.}
\label{Fig-Wollaston}
\end{figure}

A last strategy to create the two beams of the measurement area is illustrated in Fig. \ref{Fig-Wollaston}. It is very similar to that of ref.~\cite{denBoef-1991}. The initial beam (field $\mathbf{E}_{0}$) first crosses a Wollaston beam splitter, which produces 2 rays of orthogonal polarizations ($\mathbf{E}_{\base}$ and $\mathbf{E}_{\tip}$), with an $\alpha=\ang{2}$ separation angle. If the intersection point of these beams is placed at the focal point of the focusing lens L$_{0}$, the resulting beams are focused on the sensor, their optical axes being parallel. Translation of the Wollaston prism along the optical axis does not change the focalization points, but the angle of incidence of the beam on the sensor. It can thus be used to correct for a static curvature of the cantilever.

\begin{figure}
 \begin{center}
 	\includegraphics{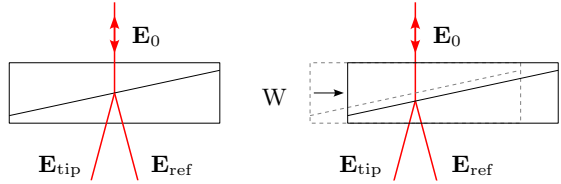}
\end{center}
\caption{Tuning the initial optical path difference: a translation of the Wollaston prism W along the separation axis modifies the optical path of the two polarized beam ($\mathbf{E}_{\base}$ and $\mathbf{E}_{\tip}$) inside the birefringent material. The small shift of the splitting point of the beam has a negligible effect in our setup.}
\label{Fig:Wollaston2}
\end{figure}

The distance between the 2 focalization points is given by the angular separation of the prism $\alpha=\ang{2}$ and the focal length $f=\SI{30}{mm}$, that is $\SI{1}{mm}$ in our setup. This distance is always greater than the length of the cantilever, so the reference beam is reflected on the chip holding the sensor in this configuration. It is thus a little bit out of focus, the typical thickness of the chip being \SI{0.4}{mm}. Anyway, the reflecting surface is very large and no light is lost, and the small divergence of the 2 beams we will eventually analyze decreases only slightly the contrast of the interferences. 

A benefit of this configuration is the possibility to tune the initial path difference between the two beams~\cite{denBoef-1991}. Indeed, a translation of the Wollaston prism along the axis of separation changes differently the optical path of the ordinary and extraordinary rays in the birefringent element (see figure \ref{Fig:Wollaston2}). For a lateral displacement $\delta x$ of our prism, we measure this dependence to be $\delta L / \lambda = \delta x / (\SI{10}{\micro m})$, where $\lambda=\SI{633}{nm}$ is the wavelength of the He-Ne laser we are using.

The three configurations of the measurement area, illustrated in Figs. \ref{Fig-bicalcite} to \ref{Fig-Wollaston}, have their own advantages and drawbacks, and we chose for every experiment the best compromise for our specific needs. The first setup (Fig. \ref{Fig-bicalcite}) is for example very stable with respect to external disturbances, the 2 beams sharing almost the same path except for the very last millimeters before the cantilever, but it is restricted to flat rectangular cantilevers longer (but ideally not too much) than \SI{140}{\micro m}. The double path arrangement (Fig. \ref{Fig-DoublePath}) can be more sensitive and adapted to curved sensors, but is more vulnerable to external vibrations and more complex to tune. The Wollaston configuration (Fig. \ref{Fig-Wollaston}) is easy to align and adapted to cantilever of any length and shape (V shaped for instance), but it is also more affected by environmental perturbations and its contrast is not as sharp. A further comparison of the Wollaston and Bi-calcite configurations, based on thermal noise measurement, is conducted in part~\ref{subsection:ThermalNoise}

\subsection{Interferometer: analysis area}

\begin{figure}
	\begin{center}	   
	\includegraphics{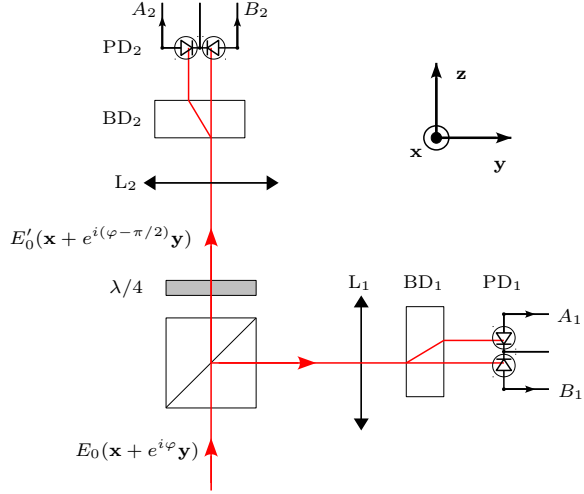}
	\end{center}
\caption{Experimental setup: analysis area. The light coming from the measurement area is split into two arms, indexed with subscript $n=1,2$. In each one, a $\SI{5}{mm}$ calcite prism (beam displacers BD$_{n}$) oriented at $\ang{45}$ with respect to the measurement birefringent component (BD$_{0}$ or W) projects the polarizations to have them interfere. The 2 beams emerging from BD$_{n}$ are focused by a plano convex lens (L$_{n}$, $f=25mm$) on the 2 segments of a 2 quadrant photodiode PD$_{n}$ to record their intensities $A_{n}$, $B_{n}$. Those are used to reconstruct $\varphi$ and thus measure the deflection $d$ of the cantilever. In the second analyzing arm ($n=2$), a quarter wave plate ($\lambda/4$) is added in order to add $\psi_{2}=-\pi/2$ to the phase shift $\varphi$. }
\label{Fig-analysis}
\end{figure}

The three setups for the measurement area share the same principle: the incident light is split in 2 with a birefringent component, and each ray is focused on the sensor. After reflection, the beams are merged into a single one and the optical path difference between its two polarizations is a linear function of the deflection of the cantilever. The analyzing area is thus the same for all the measurement configurations. It is based the quadrature phase technique we have developed~\cite{Bellon-2002-OptCom,Paolino-2007}, as illustrated on Fig. \ref{Fig-analysis}: the light is separated in two equivalent arms (indexed with subscript $n=1,2$) with a non polarizing cube beam splitter, then focused on the photodetectors by lenses L$_{n}$ (focal length \SI{25}{mm}). To record interferences, the initial polarizations are projected by calcite beam displacers (BD$_{n}$, \SI{5}{mm} thick) whose axes are at \ang{45} with respect to the optical axes of the measurement birefringent component (BD$_{0}$ or W). The two beams emerging from each calcite prism are \SI{0.5}{mm} distant, they are collected on the two segments of a 2 quadrant photodiode (\emph{UDT} Spot-2DMI). The only difference between the two arms is the addition of a quarter-wave plate in the second arm, tuned to subtract $\pi/2$ to the phase shift between the 2 polarizations.

For the following computation, we will use the configuration of Fig. \ref{Fig-bicalcite}, but it can be easily extended to the other setups with minor changes. Let us call $\mathbf{x}$ and $\mathbf{y}$ the unity vectors along the optical axes of the calcites BD$_{0}$, and $\mathbf{E}_{0}=E_{0}(\mathbf{x}+\mathbf{y})$ the electric field of the incident beam, tuned to be linearly polarized at \ang{45} with respect to $\mathbf{x}$ and $\mathbf{y}$. The total field of the beams after reflection and recombination is $\mathbf{E}_{\base}+\mathbf{E_{\tip}}=E_{0}(\mathbf{x}+e^{i \varphi}\mathbf{y})$, with $\varphi$ the phase shift between the 2 polarizations. Since the optical path difference $\delta L$ is twice the deflection $d$ of the cantilever (equation \ref{eq:deltaL}), $\varphi$ is simply
\begin{equation} \label{eq:phi-vs d}
\varphi=\frac{4\pi}{\lambda}d
\end{equation}
where $\lambda=\SI{633}{nm}$ is the wavelength of the He-Ne laser we use (Melles Griot LHP 691). The optical axes of the analyzing calcite prisms BD$_{n}$ are oriented at \ang{45} with respect to BD$_{0}$ (along  $\mathbf{x}+\mathbf{y}$ and $\mathbf{x}-\mathbf{y}$ for BD$_{2}$ for example), hence the intensities of the projected beams on the 2 quadrants of the photodiodes are easily computed as
\begin{eqnarray}
 A_{n} & = & \frac{I_0}{8} (1 + \cos(\varphi+\psi_{n})) \nonumber \\ 
 B_{n} & = & \frac{I_0}{8} (1 - \cos(\varphi+\psi_{n}))  \label{eq:intensities}
\end{eqnarray}
where subscript $n$ stand for the analyzing arm ($n=1,2$), $I_0$ is the total intensity corresponding to the light reflected by the cantilever~\cite{Note1}, and $\psi_1=0$ (first arm, without quarter wave plate) or $\psi_2=-\pi/2$ (second arm, with quarter wave plate). The photocurrents are converted to voltages by home made low noise preamplifiers directly behind the photodiodes (AD8067 operational amplifiers with $\SI{100}{k\Omega}$ retroaction and \SI{1}{MHz} bandwidth) and directly sampled using high precision digitizers (NI-PXI 5922). Using post-acquisition signal processing, we can measure for each arm the contrast function of these two signals:
\begin{equation} \label{eq:realcontrast}
C_n=\frac{A_{n} - B_{n}}{A_{n} + B_{n}}=\cos(\varphi+\psi_{n})
\end{equation}
This way, we get rid of fluctuations of laser intensity, and have a direct measurement of the cosine of the total phase shift for each arm, $\varphi+\psi_{n}$.

Let us rewrite eq. \ref{eq:realcontrast} as:
\begin{equation} \label{eq:complexcontrast}
 C=C_1+i \, C_2=\cos(\varphi)+i\sin(\varphi)=e^{i\varphi}
\end{equation}
Under this formulation, the advantage of using two analyzing arms instead of one is obvious: it allows one to have a complete determination of $\varphi$ (modulo $2\pi$). In the $(C_1,C_2)$ plane, a measurement will lay on the unit circle, its polar angle being the optical phase shift $\varphi$. The sensitivity to detect small variations in the deflection of the cantilever appears this way to be independent of the static deflection and the intrinsic optical path difference: 
\begin{equation} \label{eq:sensitivity}
\left|\frac{\d C}{\d d}\right|=\frac{4 \pi}{\lambda}
\end{equation}
No tuning of the zero is necessary. The use of crossed polarizations for the two interfering beams is a key point of this method, since it allows a post processing of the phase difference (with the quarter-wave plate) to produce the quadrature phase signals.  

Eventually, to measure the deflection $d$ of the cantilever, all what we need to do is to acquire the two contrast $C_{1}$ and $C_{2}$, and reconstruct $d$ with standard digital data processing tools: $\varphi=4 \pi d / \lambda$ can be computed as
\begin{equation} \label{eq:arctan}
\varphi = \arg(C) = \arctan(C_2/C_1)
\end{equation}
Unwrapping of the phase $\varphi$ to avoid discontinuities every $2 \pi$ can be necessary for large amplitudes of deflection. This can easily be performed as long as the sampling frequency $f_{s}$ is sufficient: the distance between successive points should be smaller than $\pi$, which translates of into $f_{s}>4\dot{d}/\lambda$ (where $\dot{d}$ is the time derivative of $d$).

\begin{figure}[t]
 \begin{center}
	\includegraphics{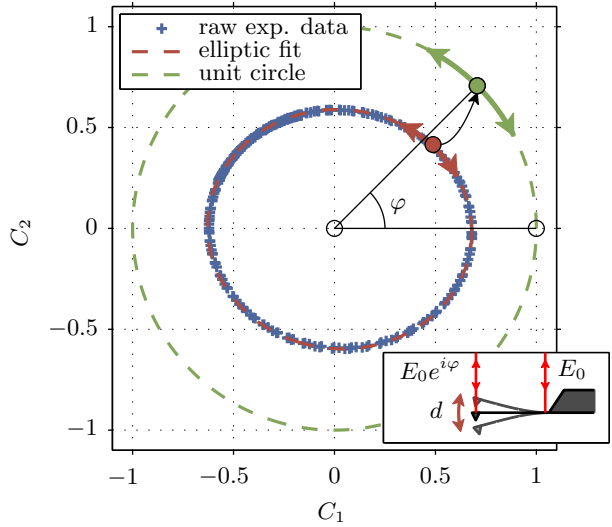}
\end{center}
\caption{Calibration of the complex contrast: we drive the cantilever close to its first resonance with an amplitude of deflection $d$ of the order of the laser wavelength $\lambda$. The experimental data follow an ellipse in the ($\widetilde{C}_{1},\widetilde{C}_{2}$) plane. After a simple fit of this ellipse with equation~\ref{eq:ellipsecontrast}, it is straightforward to project the measurement point on the unit circle, and identify its polar angle with the optical phase shift $\varphi$. Equation \ref{eq:phi-vs d} eventually leads to a calibrated measurement of deflection $d$.}
\label{Fig:Cstar}
\end{figure}

In the experimental realization, the unavoidable imperfections of optical components and of their alignment will lead the complex contrast to lay in reality on a tilted ellipse with axes smaller than one, instead of the unit circle: we can generically write
\begin{equation} \label{eq:ellipsecontrast}
\widetilde{C}=\widetilde{C}_1+i \, \widetilde{C}_2=\mathcal{C}_{1}\cos(\varphi)+c_{1}+i \left(\mathcal{C}_{2}\sin(\varphi+\psi)+c_{2}\right)
\end{equation}
where over tilde contrasts represent the measured values (by opposition to the ideal values of equation \ref{eq:complexcontrast}), and $\mathcal{C}_{n}<1$ are the contrast amplitudes in each arm, $c_{n}$ the contrast offsets, and $\psi$ a residual mismatch to perfect quadrature~\cite{Bellon-2002-OptCom}. These 5 parameters can easily be extracted from a calibration of the interferometer: we excite a small amplitude oscillation at resonance of the free cantilever with a piezo, and get the parameters from a generic fit of the recorded ellipse (see figure \ref{Fig:Cstar}). In the Wollaston configuration, we can also produce the ellipse with a lateral driving of the birefringent prism, allowing us to calibrate the system with no oscillation of the cantilever. The results of these 2 methods in this case are in perfect agreement. Once this elliptic fit done, raw measurements ($\widetilde{C}$) can easily be post processed and projected on the unit circle ($C$), to extract the actual deflection~\cite{Bellon-2002-OptCom}.

\begin{figure}
 \begin{center}
	\includegraphics{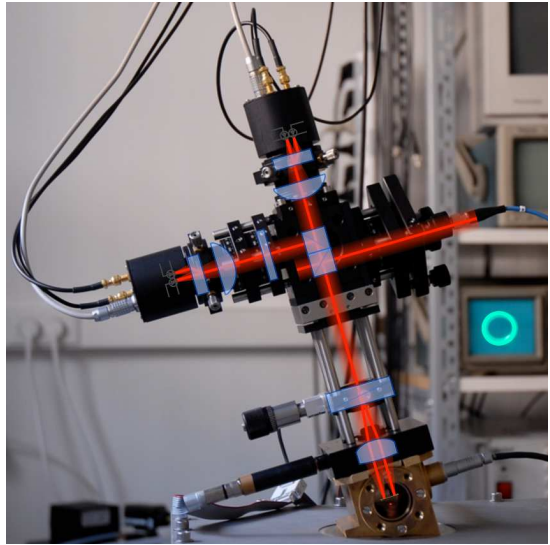}
\end{center}
\caption{Photograph of our experimental realization of the quadrature phase interferometer for AFM measurement with superposed optical scheme (Wollaston configuration). The laser light is fed into the setup with an optical fiber (blue cord on the right of the image). In the background, an oscilloscope in XY mode displays the elliptic trace of the outputs during the procedure of calibration of the complex contrast.}
\label{Fig:AFMphoto}
\end{figure}

Let us summarize the main points of our interferometric setup. The input laser light is split into a reference beam, directed on the base of the cantilever, and a sensing beam, focused on top of the tip of the sensor. The phase shift $\varphi$ is a linear function of the deflection $d$ (equation \ref{eq:phi-vs d}). After reflection, the two rays of crossed polarization are recombined and processed with a quadrature phase technique: two output signals $\widetilde{C}_1$ and $\widetilde{C}_2$ are produced, with $\widetilde{C}_{1}\sim \cos \varphi$, and $\widetilde{C}_{2}\sim \sin \varphi$ (equation \ref{eq:ellipsecontrast}). Post-processing of these signals leads to a precise, calibrated and virtually unbounded value of the deflection $d$. Figure~\ref{Fig:AFMphoto} illustrate our experimental realization by a photograph of the setup with superposed optical scheme.

The strategy proposed in this paper (post-processing of the signals) is intended and adapted to force spectroscopy. For AFM imaging however, a real-time deflection signal is needed for the feedback loop. For this purpose, we also performed an analog implementation of equation \ref{eq:realcontrast} (using AD734 integrated circuit for division). Using the Wollaston prism, we can tune the optical phase close to 0 at the chosen deflection set-point $d_{sp}$ (see figure \ref{Fig:Wollaston2}), therefore the signal $\tilde{C}_{2}\sim \sin(4 \pi (d -d_{sp}) / \lambda)$ can be used as the error signal of the feedback loop. Other strategies adapted to a large deflection range, random initial phase, or dynamic AFM are also possible~\cite{Aguilar-inpreparation}.

\section{Performances of our experimental realization}

We now present the main performances in terms of stability and precision of the two main configurations  (bi-calcite and Wollaston). The double-path technique, due to its complexity of tuning, has mainly been used to deal with curved cantilevers, and was not systematically characterized.

\subsection{Long term stability}

\begin{figure}
 \begin{center}
	\includegraphics{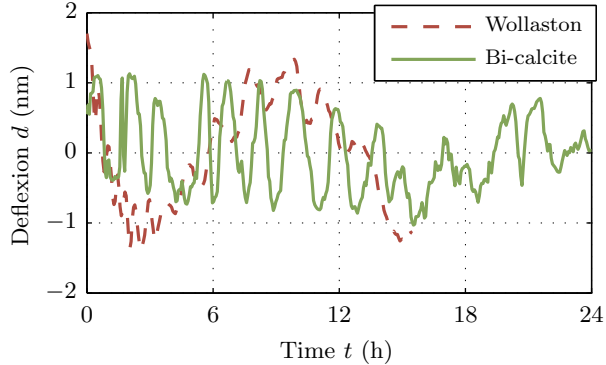}
\end{center}
\caption{Long term stability of the interferometric measurement: after calibration, an acquisition of the deflection of a cantilever at rest is performed. For both the bi-calcite and the Wollaston configuration, the drift after several hours is at most $\SI{3}{nm}$.}
\label{Fig:Stability}
\end{figure}

We present in figure \ref{Fig:Stability} the evolution over several hours of the deflection measured on a cantilever at rest, far from any sample. Slow environmental changes (temperature drift...) have little influence on the measurement: the deflection is stable within a $\SI{\pm 1.5}{nm}$ fluctuation range, for both sensing strategies. This long term stability is excellent: for most applications, there is no need for frequent calibrations of the interferometer. 

\subsection{Thermal noise measurement} \label{subsection:ThermalNoise}

\begin{figure}
 \begin{center}
	\includegraphics{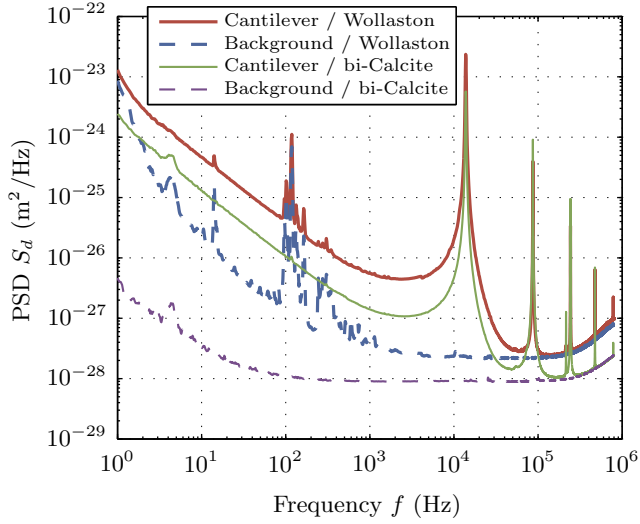}
\end{center}
\caption{Power spectrum density of deflection measured in the Wollaston and bi-calcite configurations of a golden coated cantilever in air and on a rigid mirror (background noise). The raw data plotted here demonstrate the high precision of the interferometer: the background noise is as small as $\SI{E-14}{m/\sqrt{Hz}}$ for the bi-calcite set-up, on a wide frequency range. This is only $\SI{20}{\%}$ higher than the unavoidable shot noise due to the photodiodes, and a small increase can be noticed at low and high frequencies due to higher conditioning electronics noise. In the Wollaston configuration, the base line noise is slightly higher at $\SI{1.4E-14}{m/\sqrt{Hz}}$ (again within $\SI{20}{\%}$ of the shot noise limit), and degrades faster at low frequency: the setup is more sensitive to external perturbations as the reference and sensing laser beams are separated on a longer distance. In both configuration however, the thermal noise of the cantilever is clearly visible, with 5 perfectly defined flexural modes in this $\SI{1}{MHz}$ frequency range. The $1/f$ like behavior at low frequency, 2 orders of magnitude larger than the background noise, is studied in references~\cite{Paolino-2009-Nanotech,Li-2012}.}
\label{Fig:BiCvsWraw}
\end{figure}

This section on the thermal noise of a cantilever is just a quick illustration of the high precision of our interferometric measurement of the deflection, an extensive study on this topic can be found in references~\cite{Bellon-2008,Paolino-2009-JAP,Paolino-2009-Nanotech,Li-2012,Laurent-2013}.

We use our setup to record the equilibrium fluctuations of a cantilever at rest, far from any sample. Thermal noise acts as a white noise random forcing, exciting all resonant modes of the microscopic mechanical beam: as shown on figure \ref{Fig:BiCvsWraw}, the first 5 structural resonances are visible on the Power Spectrum Density (PSD) $S_{d}$ in the $\SI{1}{MHz}$ range in frequency explored here. The data was acquired on a gold coated cantilever \emph{Budget Sensors} Cont-GB in air. The cleanness of the curves demonstrates the sensitivity of the detection method, down to low frequencies. The instrument noise for the bi-calcite configuration, recorded by replacing the cantilever by a rigid mirror while keeping intensities on the photodiodes unchanged, decreases from $\SI{E-13}{m/\sqrt{Hz}}$ at $\SI{1}{Hz}$ down to $\SI{E-14}{m/\sqrt{Hz}}$ for frequencies $f$ above $\SI{100}{Hz}$, before raising gently again above $\SI{300}{kHz}$. The main contribution to this floor noise is the unavoidable shot noise due to the photodiodes. We can estimate this contribution from the light intensities recorded at each photodetector: the performance is close to optimal conditions, only $20\%$ above the shot noise limit for a laser power around $\SI{100}{\micro W}$ per sensing beam. On top of this white noise, the conditioning electronic noise is added (raising the noise at low and high frequency), and some environmental vibration can be seen (bump in the $\SI{10}{Hz}$ region). 

The Wollaston configuration is more sensitive to external perturbations, especially at low frequency where its background noise is not as low. Many more vibrations peaks due to environmental vibrations can be noted below $\SI{1}{kHz}$. However, the mechanical noise intrinsic to the cantilever is still almost everywhere much higher than this detection limit at low frequency. For $f$ in the $\SI{1}{kHz}-\SI{300}{kHz}$ range, the background fluctuations reach again the shot noise limit, at $\SI{1.4E-14}{m/\sqrt{Hz}}$. 

\begin{figure}[t]
 \begin{center}
	\includegraphics{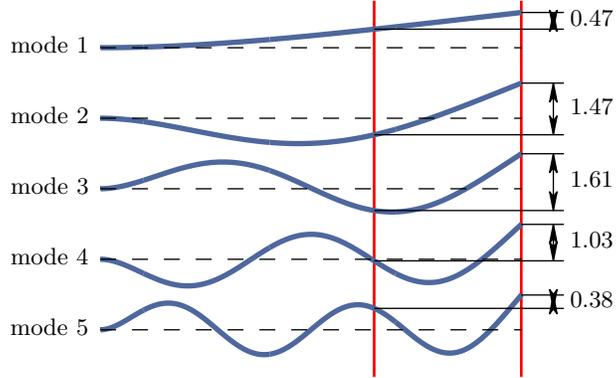}
\end{center}
\caption{In the bi-calcite configuration, the distance between the two laser spots is smaller than the length of the cantilever, thus the measured deflection does not correspond to the actual value. The correction to consider is mode dependent: according to the position of the spots with respect to the nodes and antinodes of the spatial shape, it can be smaller or larger than one. Using an Euler Bernoulli model for the mechanical beam, we estimate the coefficient noted on the figure for the measured amplitude normalized to the deflection of the free end for the 5 first modes.}
\label{Fig:BiCModeCoef}
\end{figure}

\begin{figure}[hb]
 \begin{center}
	\includegraphics{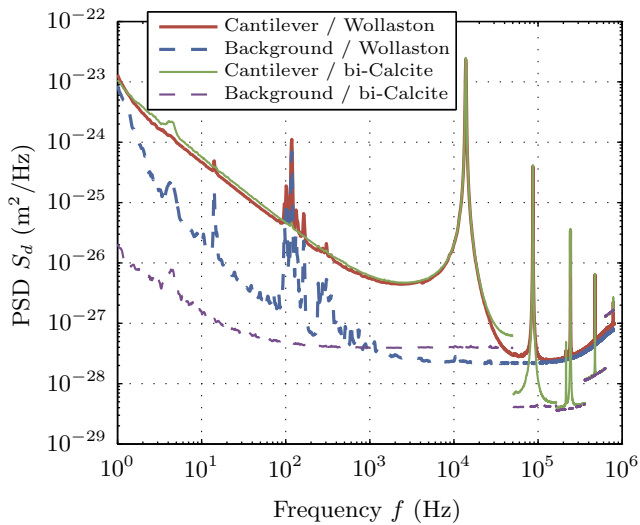}
\end{center}
\caption{Power spectrum density of deflection measured in the Wollaston and bi-calcite configurations of a golden coated cantilever and on a rigid mirror (background noise). The spectrums measured in the bi-calcite configuration have been corrected to take into account the real position of the laser beams on the cantilever: since the cantilever is longer (\SI{450}{\micro m}) than the beam separation (\SI{140}{\micro m}), the reference point is not static. A mode dependent correction has been considered.}
\label{Fig:BiCvsW_modecorrection}
\end{figure}

We note a discrepancy between the two measurement methods: the amplitude of the spectrums are not equal, with a factor depending on the mode considered. Indeed, as the cantilever is $\SI{450}{\micro m}$ long, we have to correct the output of the bi-calcite configuration: the spots are only $\SI{141}{\micro m}$ distant, so the actual deflection is not equal to the interferometer output. The multiplicative factor to consider is mode dependent: as illustrated on figure \ref{Fig:BiCModeCoef}, the coefficient is about $2$ for the first mode, but can be smaller than one for higher order modes. We therefore correct the power spectrums of figure \ref{Fig:BiCvsWraw} for the bi-calcite configuration considering those factors (we switch between coefficient at half way between resonance frequencies). The final curve presented in figure \ref{Fig:BiCvsW_modecorrection} matches closely the measurement with the Wollaston setup, except when the fluctuations are close to the background noise (between resonances).

This mode dependent correction is anyway a complex procedure, which has to be tuned for every specific geometry. The background noise, in absolute better for bi-calcite configuration, can eventually get worse than that of the Wollaston configuration. Moreover, in between resonances, where the noise of the modes are of equivalent amplitude, further assumptions would be required to extract information from the measured fluctuation. Finally, the study of torsional modes is difficult with this setup as it requires a perfect alignment between the cantilever and the two laser spots. On figure \ref{Fig:BiCvsWraw} for instance, a peak of small amplitude can be noticed just before the third flexural mode resonance, only in the bi-calcite configuration: we couldn't perfectly center both sensing beams, and the first torsional mode is visible on the spectrum. Unless the cantilever length is close to the laser spots distance, the bi-calcite configuration is thus not adapted to the simultaneous study of several modes, nor to the absolute characterization of the cantilever noise. However, its excellent performance at low frequency is ideal to study the low frequency behavior of thermal fluctuations, and this configuration is sometimes preferable, in the study of the $1/f$ noise at low frequency for example~\cite{Paolino-2009-Nanotech,Li-2012}.

\subsection{Linearity of the sensor}

\begin{figure}
 \begin{center}
	\includegraphics{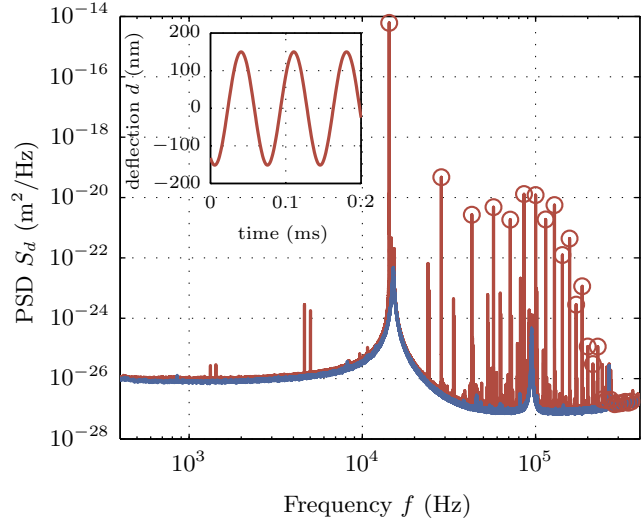}
\end{center}
\caption{Power spectrum density of deflection measured in the Wollaston configuration of an Al coated cantilever with (red) and without (blue) an external driving of $\SI{300}{nm}$ close to the resonant frequency. Harmonics of the driving frequency are tagged with open circles. The total harmonic distortion is only $\SI{1.5e-5}{}$, demonstrating the high linearity of the sensor on this large input range. The thermal noise spectrum is perfectly defined even during the oscillation, illustrating the high dynamic input range of the instrument. Inset: time trace of the deflection on a few oscillations.}
\label{Fig:HD}
\end{figure}

To test the linearity of our interferometric method on a wide input range, we drive a cantilever close to its first resonance with a small piezo oscillation of its holder. The driving function generator has an excellent harmonicity, and even if the piezo response is not perfectly linear, thanks to the resonant behavior of the cantilever, the resulting deflection should be almost purely sinusoidal (except for the thermal noise contribution). The amplitude of the oscillation is chosen so that the complex contrast explore the full calibration ellipse, that is to say that the deflection is sinusoidal with peak-peak amplitude of $\lambda/2\sim\SI{300}{nm}$, as shown in the inset of figure \ref{Fig:HD}.

The PSD of the measured deflection with and without the driving reported in figure \ref{Fig:HD} demonstrate the high dynamic range of your measurement: the thermal noise spectrum, of RMS amplitude $\SI{0.13}{nm}$ is perfectly defined and unchanged by the $\SI{300}{nm}$ oscillation. Harmonics of the excitation signal are visible, but their amplitudes are orders of magnitude lower than that of the fundamental. The total harmonic distortion (THD) evaluated on this spectra is only $\SI{1.5e-5}{}$. We tested oscillation amplitudes up to $\SI{3}{\micro m}$ retaining similar THDs around $\SI{e-5}{}$, demonstrating the high linearity of the sensor on this large input range. We used this property for instance to evaluate thermal noise spectra during force curves on soft carbon nanotubes, retaining all useful spectral resolution while measuring deflection up to a few hundred nm~\cite{Buchoux-2011}.

\subsection{Comparison with other devices}

To conclude this section on our deflection measurement setup, we compare its performance to that of commercial AFMs and other interferometric detection systems in the literature.

Most commercial AFMs today use an optical lever configuration~\cite{Meyer-1988}: a laser beam is focused on the cantilever, and the reflected light is collected on a 4 quadrants photodiode. A deflection changes the position of the spot on the detector, thus the relative intensities of each segment proportionally to the slope of the illuminated part of the cantilever. This method is very simple to implement, and achieves a high precision: using dedicated cantilever geometries, it can even be as sensitive as an interferometric approach~\cite{Putman-1992,Gustafsson-1994}. However, for common geometries and commercially available cantilevers, the shot-noise limit of the optical lever technique is around ten time higher than our interferometric approach~\cite{Note2}: the background noise for the cantilever of figure \ref{Fig:BiCvsWraw} is at best $\SI{E-13}{m/\sqrt{Hz}}$. Moreover, the electronic and environmental noise are most of the time not as low as in our experiment, making low frequency measurements of thermal noise impossible with commercial devices. Lastly, a calibration step is required to convert the output of the photosensor into the actual deflection of the cantilever: during a contact with a hard surface, the sample is moved vertically of a known distance, leading to the calibration constant. This step can be undesirable, to prevent any alteration of the tip during the hard contact for instance, and can be avoided with interferometric approaches. Moreover, this calibration procedure is accurate only for static deflections, and the correction to consider depends non trivially on the mode considered at higher frequency~\cite{Schaffer-2005}.

\begin{figure}
 \begin{center}
	\includegraphics{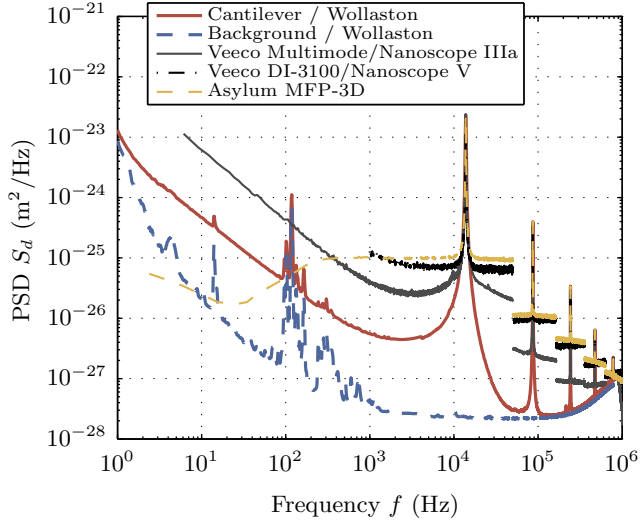}
\end{center}
\caption{Power spectrum density of deflection measured in the Wollaston configuration and on 3 commercial devices. The same golden coated cantilever is used, except for the background noise of our setup which is acquired on a rigid mirror. The spectrums measured with commercial devices have been corrected to convert the angular measurement into a deflection measurement. The correction is mode dependent, and tuned so that the amplitude of each resonance matches the calibrated spectrum.}
\label{Fig:CommercialSpectrums}
\end{figure}

We present in figure \ref{Fig:CommercialSpectrums} a comparison of noise measurements with our setup (Wollaston configuration) and three commercial devices, on the same cantilever. A mode dependent multiplicative factor has been applied to the three spectrums acquired on commercial apparatuses, so that the amplitude of each resonance matches our calibrated measurement. The background noise of the three detection systems is dominant everywhere except at the resonances, and is at least 10 time worse (in $\SI{}{m/\sqrt{Hz}}$ units) than in our interferometer, though the laser spot intensity is 10 times higher (around $\SI{1.1}{mW}$ in the 3 devices). Low frequency information on the mechanical thermal noise is not accessible:
\begin{itemize}
\item for the 2 Veeco devices, the electronic $1/f$ noise is dominant. The spectrum below $\SI{1}{kHz}$ is not accessible in the Nanoscope V due to software limitation. It was acquired using a Signal Access Module and independent acquisition cards on the Nanoscope IIIa.
\item for the Asylum AFM, the low frequency noise is processed with a high pass filter, cutting the electronic $1/f$ noise but also any physical information about the system.
\end{itemize}
Let us stress anyway that the high sensitivity of 4 quadrant detectors is more than sufficient for imaging purposes. Its simplicity drove its election in most commercial systems. However, in thermal noise study or force spectroscopy, the higher resolution and intrinsic calibration of our approach can lead to original results.

Before the optical lever technique became dominant, a bloom of various approaches to read the cantilever deflection followed the invention of the AFM by Binnig, Quate and Gerber in 1986~\cite{Binnig-1986}. Interferometric methods were and still are the most sensitive methods, and exist nowadays as home built systems in a few laboratories. We will focus only on homodyne interferometers, since heterodyne setups~\cite{Martin-1987} are usually limited to dynamic operation and cannot access the static information which is of interest for force spectroscopy applications. They can be categorized into two main groups:
\begin{itemize}
\item \emph{Fabry-P\'{e}rot like interferometers}. In its most simple realization, the fiber interferometer, the resonant cavity is composed by the cantilever and the cleaved end of an optical fiber~\cite{Rugar-1988,Rugar-1989}. The finesse of the cavity in this case is close to 1, as multiple reflections are killed by the low reflectivity of the fiber end and the divergence of the optical beam reflected by the cantilever. The output of the interferometer is then a sinusoidal function of the deflection. Background noise down to $\SI{5e-14}{m/\sqrt{Hz}}$ have been reported~\cite{Rugar-1989} for this arrangement for incident light power around $\SI{100}{\micro W}$. To increase the finesse of the interferometer and thus its sensitivity, a reflective coating covering the fiber end has been proposed in references~\cite{Mulhern-1991,Mamin-2001,Oral-2003,Patil-2005}. A drawback of this design is the coupling between deflection and light intensity in the cavity: photothermal and radiation pressure bending the cantilever are function of the light intensity which is linked to the deflection. This can result into a self oscillating as well as a self cooling cantilever~\cite{Metzger-2004,Vogel-2003}, eventually complicating the data analysis. Hoogenboom and coworkers~\cite{Hoogenboom-2005,Hoogenboom-2008}, in a convergent geometry where the cantilever is at the focal point of a concave semi-reflecting lens, reached a finesse of 20. The output is in this case highly non linear, but for deflection much smaller than the wavelength around the optimum of sensitivity, a background noise of $\SI{E-15}{m/\sqrt{Hz}}$ at high frequency ($\SI{1}{MHz}$ and higher) can be reached for an incident light power of $\SI{1}{mW}$. The noise at lower frequencies (less than $\SI{250}{kHz}$) is however at best comparable to the performance of our system.
\item \emph{Differential interferometers}. These interferometers~\cite{Schonenberger-1989,denBoef-1991,Anselmetti-1992,Cunningham-1994,Goto-1995} have inspired this work. They use birefringent elements to split an incident light into two beams of crossed polarization which are both focused on the cantilever (or on a reference mirror strongly connected to it~\cite{Cunningham-1994,Goto-1995}). Those systems have shot noise limited performances, and as such are as good as our realization: the baseline for the background noise reached in these experiments is around $\SI{E-14}{m/\sqrt{Hz}}$ for comparable incident light power. 
\end{itemize}

Our experimental device, described in details in this chapter, thus reaches one of the best precision one can hope for only $\SI{100}{\micro W}$ incident light power on the cantilever. It presents moreover a significant advantage over every other configuration: the deflection range for which this performance is available is very wide, up to a few $\SI{}{\micro m}$. The convergent Fabry-P\'{e}rot configuration~\cite{Hoogenboom-2005} for example, reaching the best absolute performance, implies a cavity finesse of 20, limiting the linear deflection range to $\SI{20}{nm}$ at best. With other interferometer design, the typical range available is smaller than $\lambda/4$, thus around $\SI{100}{nm}$. High precision optical lever set-ups~\cite{Fukuma-2009} also present a limitation around $\SI{100}{nm}$ for their linear deflection range (and require an additional calibration step, as already mentioned).

\section{Conclusion}

Let us summarize the main properties of our interferometric measurement of AFM cantilever deflection:
\begin{itemize}
\item A very high resolution, down to $\SI{E-14}{m/\sqrt{Hz}}$, on a very wide spectral range, equaling or out-performing the best results reported in the literature for equivalent light power, especially at low frequency.
\item A huge deflection range, up to several $\SI{}{\micro m}$, much larger than any other high precision systems.
\item An intrinsically calibrated measurement, thanks to the interferometric approach.
\item A very stable sensor, with less than $\SI{3}{nm}$ drift over $\SI{24}{h}$.
\end{itemize}

\begin{figure}
 \begin{center}
	\includegraphics{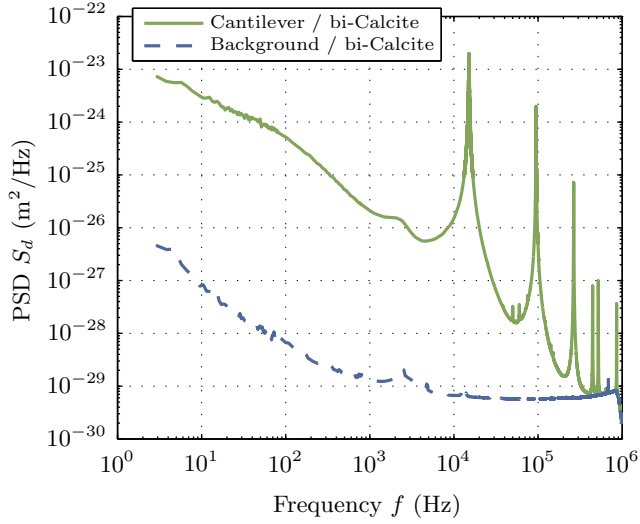}
\end{center}
\caption{Power spectrum density of deflection measured in the calcite configuration of an Al coated cantilever and on a rigid mirror (background noise) with a green DPSS laser: the light power is 100 times larger than with the He-Ne Laser, resulting in a lower background noise, down to $\SI{6e-30}{m^{2}/Hz}$.}
\label{Fig:C532}
\end{figure}

Since the ultimate precision of our interferometer in shot noise limited, the natural next step is to try to increase the laser power to enhance even further the resolution. We recently acquired a diode pumped solid state laser (Spectra Physics Excelsior 532 Single Mode), spectral linewidth lower than $\SI{10}{MHz}$, with an output power up to $\SI{100}{mW}$ at $\SI{532}{nm}$. The first spectrums acquired are full of promises, as shown on figure \ref{Fig:C532}: with a $\SI{12}{mW}$ light power per sensing beam, we were able to lower the background noise down to $\SI{2.5e-15}{m/\sqrt{Hz}}$ on an aluminum coated cantilever. If the effect of such a large intensities (such as photothermal effects) are still to explore, this experiment demonstrates that even sharper details could be detected with this set-up, pushing the limit of detection noise more than 2 orders of magnitude lower than that of the best commercial systems.

\begin{acknowledgments}
We thank F. Vittoz and F. Ropars for technical support, and N. Garnier, S. Joubaud, S. Ciliberto and A. Petrosyan for stimulating discussions. This work has been partially supported by contract ANR-05-BLAN-0105-01 of the Agence Nationale de la Recherche in France, the F\'ed\'eration de Physique A.M. Amp\`ere in Lyon, the MECESUP2 scholarship program in Chile, and the ERC project Outeflucop (for the latest developments with the DPSS Laser).
\end{acknowledgments}


\end{document}